\documentclass[preprint,showpacs,prl,aps]{revtex4}

\usepackage{graphicx}

\begin{document}

\title{Simple external cavity diode laser}

\author{O.~I.~Permyakova}

\affiliation{Institute of Semiconductor Physics,
Russian Academy of Sciences, 630090 Novosibirsk, Russia}

\author{A.V.~Yakovlev}

\author{P.L.~Chapovsky\thanks{E-mail: chapovsky@iae.nsk.su}}
\email[E-mail: ]{chapovsky@iae.nsk.su}

\affiliation{Institute of Automation and Electrometry,
          Russian Academy of Sciences, 630090 Novosibirsk, Russia}
         
\date{\today}

\begin{abstract}
A semiconductor diode laser having a modified Littrow external
resonator is described. An additional output coupling mirror in a V-shape
configuration of the resonator makes the system more efficient and convenient
to operate.
\end{abstract}

\pacs{42.60.P; 42.55.P; 42.62.F}

\maketitle

\section{Introduction}

Well-known achievements in atomic physics of the last two decades became
possible due to the progress in the semiconductor diode lasers. 
External cavity diode lasers (ECDL) are the most important in this context. 
They are able to produce powerful coherent radiation in the vicinity of
many important atomic resonances. Details of various diode laser designs and
laser performance are described in a number of papers, see, e.g.,  
Refs.~\cite{Wieman91RSI,MacAdam92AJP,Maki93OC,Corwin98AO,Yashchuk00RSI}.  

\begin{figure}[htb]
\includegraphics[width=14cm]{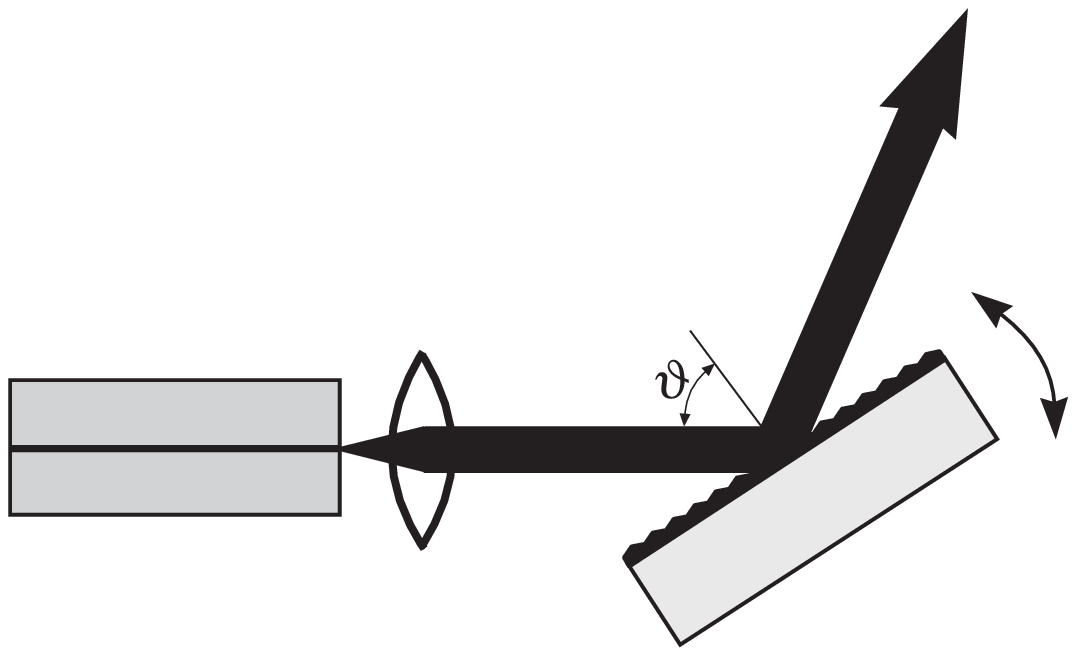}
\caption{External cavity in the Littrow configuration.}
\label{fig1}
\end{figure}

Most popular arrangement of an external cavity of a diode laser is
based on the grating feedback in the Littrow configuration, Fig.~\ref{fig1}.
In such resonator, the grating reflects back to the diode the
radiation in non-zero diffraction order. The laser output coupling is 
performed through zero diffraction order of the grating.
Although such resonator configuration looks the simplest possible 
there are some inconveniences in its practical realization. In the present paper
we describe a modified diode laser resonator that is more
efficient and convenient. This approach to the laser wavelength selection
by grating is based on our previous experience with dye lasers
\cite{Ishchenko80}. Another source of helpful ideas for this work was an 
excellent monograph \cite{Malyshev79} devoted to the classical spectral instruments.

\section{Littrow resonator}

Performance of the laser having Littrow resonator (Fig.~\ref{fig1})
is determined by the
three major factors: grating angular dispersion, grating reflection
coefficient and output coupling coefficient.  
The grating formula for the Littrow configuration, 
\begin{equation}
     2d\sin\theta=m\lambda,
\label{grating}
\end{equation}
allows to deduce the angular dispersion, $D$, of the grating,
\begin{equation}
     D \equiv \frac{d\theta}{d\lambda} = \frac{m}{2d\cos\theta}.
\label{D}
\end{equation}
Here, $d$ is the distance between two neighbouring groves of the grating, 
$m$ is the diffraction order. From this 
expression one can conclude that the angular dispersion, $D$, is growing if
$d$ is decreasing. It motivates ones to use in most designs of the laser
external cavities  the gratings having large density of groves. 

One can modify the expression (\ref{D}) for $D$ by substituting the diffraction order $m$
from the grating formula (\ref{grating}) and obtain,
\begin{equation}
     D = \lambda^{-1}\tan\theta.
\label{D1}
\end{equation}
This equation shows that the angular dispersion of the grating and
consequently the frequency resolution do not 
depend on the density of groves if one works at predefined 
diffraction angle $\theta$. This conclusion allows to make an 
important simplification of the design by using
cheaper and better quality gratings having low density of groves. 

Another conclusion from Eq.~(\ref{D1}) is that the angular dispersion
is larger at larger diffraction angles $\theta$. The drawback of working at
large $\theta$ is that for an ordinary flat grating
the reflection coefficient decreases at large $\theta$. One knows (see, e.g., 
\cite{Malyshev79}) that
the reflection at large diffraction angles can be increased by using the
gratings having the groves of triangular shape that increases the grating
blazing angle. Consequently, the gratings having low density of 
groves have an additional advantage because for such gratings a modern technology 
gives better quality of groves. 

\section{V-shape resonator}  

Another important parameter determined laser performance is the magnitude of the 
resonator output coupling. Optimal output coupling is a complicated 
function of the laser medium gain, resonator losses and
power saturation mechanism. In practice, one uses to try
a few output couplers to find out the right one for better laser performance. 
For the Littrow resonator in Fig.~\ref{fig1} the output coupling 
is determined by the properties of particular 
grating and cannot be changed by will. 

\begin{figure}[htb]
\includegraphics[width=14cm]{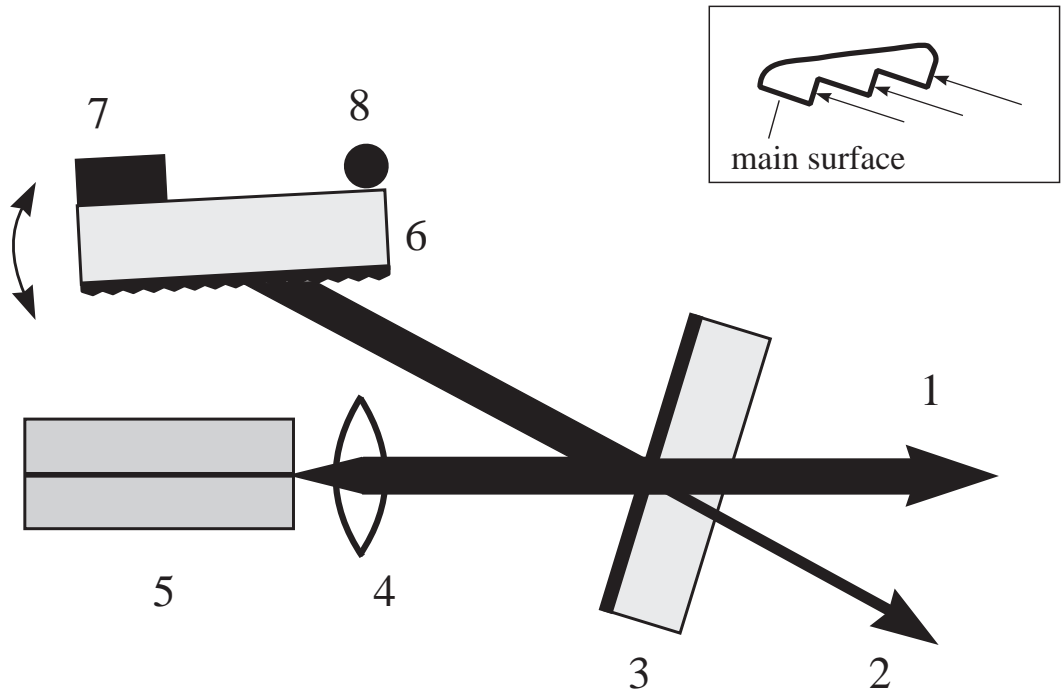}
\caption{V-shape external cavity of the diode laser. 1 - main beam; 2 - second beam;
3 - output coupling mirror; 4 - collimator; 5 - laser diode; 6 - grating; 7 - 
piezo transducer; 8 - axis of the grating rotation.}
\label{fig2}
\end{figure}

One way to solve this problem is to use an extra reflector inside 
a Littrow resonator \cite{Wieman91RSI}.  
We propose here to use an output coupling mirror in a V-shape resonator,
Fig.~\ref{fig2}.   

The grating used in this work had 600 groves/mm and the blazing angle
equal $\simeq30^0$ specified by the manufacturer. This blazing angle is
determined by the orientation of the main surface of the grove 
(see insert in Fig.~\ref{fig2}).
This surface of the grove would allow us to work at the diffraction
angles $\theta\simeq30^0$. 
Technology of the grating production is such that the groves have 
a triangular shape and one can use another surface of the 
grove to work at much larger blazing angle \cite{Malyshev79}.
In our case it allows us to work at the blazing angle, $\theta=69^0$,
that corresponds to the 4-th diffraction order for $\lambda=0.78~\mu$.
Reflection coefficient of our grating in the 4-th order was $\simeq 60\%$. The reflection 
coefficient in the 0-th order at $\theta=69^0$ appeared to be very low, 
$\sim 1\%$. In fact, this
circumstance has motivated us to modify the usual Littrow
resonator. 

We have used standard Mitsubishi laser diode, ML6XX24 series,  
without a special antireflection coating on the
crystal facet. Concequently, the laser resonator consists of the two,
one is formed by the diode crystal itself and another is formed by the 
external resonator.
Fine tuning of the laser frequency needs in this case simultaneous tuning of these
two resonators to avoid nonlinearities due to the frequency pulling effect. 
It was performed in our laser by synchronous linear modulation
of the diode current and the grating tilting by piezo transducer. The
length of the external resonator was equal to 3.5~cm. 

Stable laser characteristics demands rather precise temperature stabilization
of the laser diode. It was done in the setup with the help of Peltier element
and feedback electronics. The temperature stability was
better than 1~mK. The external laser resonator was not temperature stabilized 
actively. Instead, it was placed on aluminium plate that had good thermal 
contact with the massive metal optical table. 

\section{Laser performance}

The specified wavelength of the laser diode was 785~nm at 25~C. 
In the experiments with the rubidium vapour the laser diode 
was kept at 14~C and its own frequency at such temperature was $\simeq 781$~nm.
Power of the laser diode itself (without an external resonator) was 35~mW
at the electrical current through the  diode equal 72~mA and the temperature
equal 14~C.

The laser produces two beams (Fig.~\ref{fig2}). Direction of the main beam
does not depend on the grating and mirror orientation.  Direction of the 
second beam changes if the laser wavelength is tuned like it does in 
the Littrow resonator.
Power of the external cavity diode laser having the flat output coupling mirror 
$T=45\%$ was 18~mW, combined from 14~mW of the main output beam and 4~mW 
of the second beam. The output coupling mirror having
$T\simeq60\%$ gives 21~mW in the main beam and 3~mW in the second beam. Thus, 
as one would expect, an adjustment of the output coupling mirror is 
significant for the laser performance. For the V-shape resonator such
adjustment appears to be rather simple procedure.

\begin{figure}[htb]
\includegraphics[width=14cm]{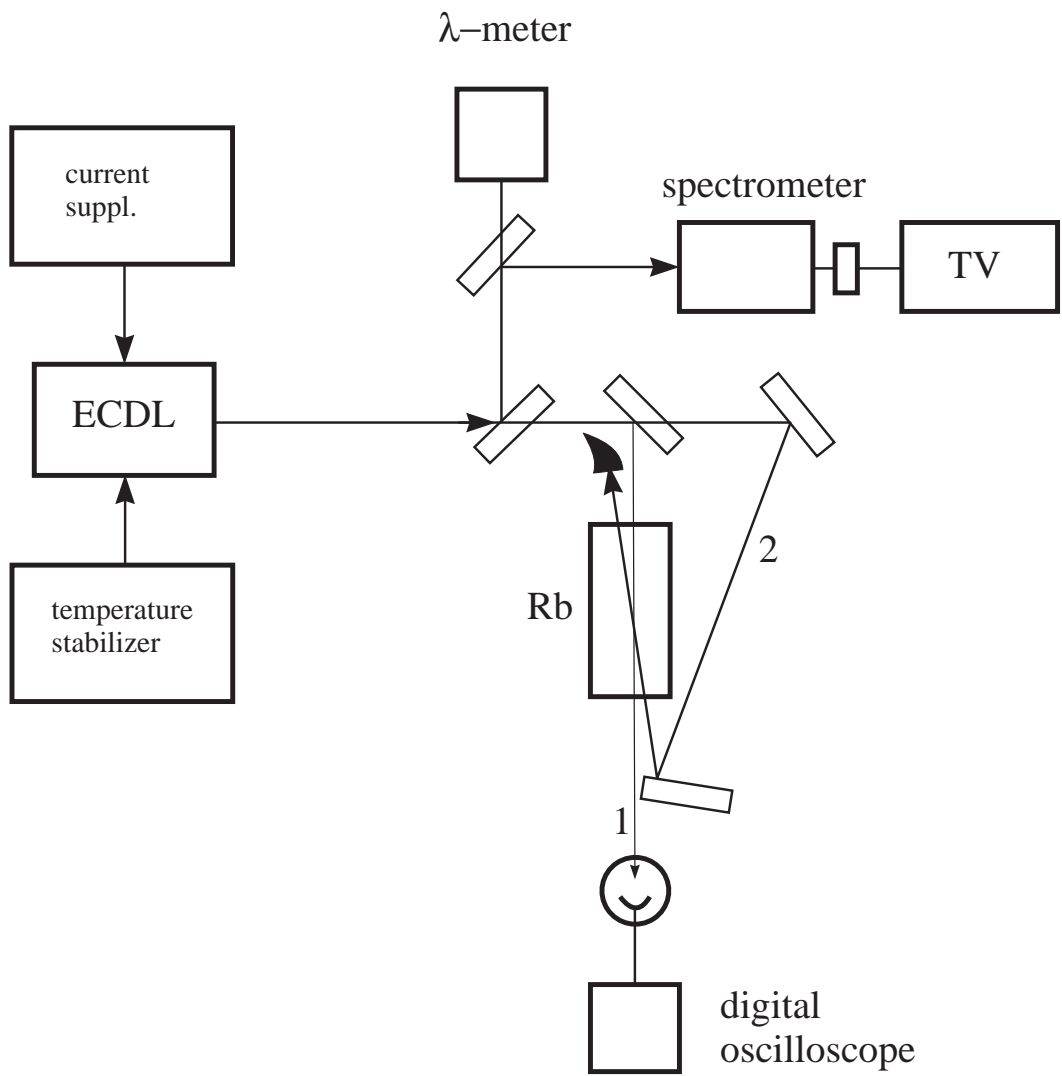}
\caption{Setup. 1 - weak beam; 2 - strong beam.}
\label{fig3}
\end{figure}

The laser frequency characteristics were measured using the setup shown in Fig.~\ref{fig3}. 
The wavelength of the laser radiation was controlled by the spectrograph having 
the dispersion equal 0.36~nm/cm on the TV screen and precise $\lambda$--meter having 
sensitivity equal 50~MHz. Rough frequency tuning of the laser radiation was 
performed  by rotating the mirror that was placed on the tuning head.
 For the output coupling mirror having
$T=45\%$ the laser wavelength was tuned between 775-785~nm at a constant 
diode temperature equal
14~C. The range of the laser frequency tuning decreases if one
uses output coupling mirror having larger transmission.

\begin{figure}[htb]
\includegraphics[width=14cm]{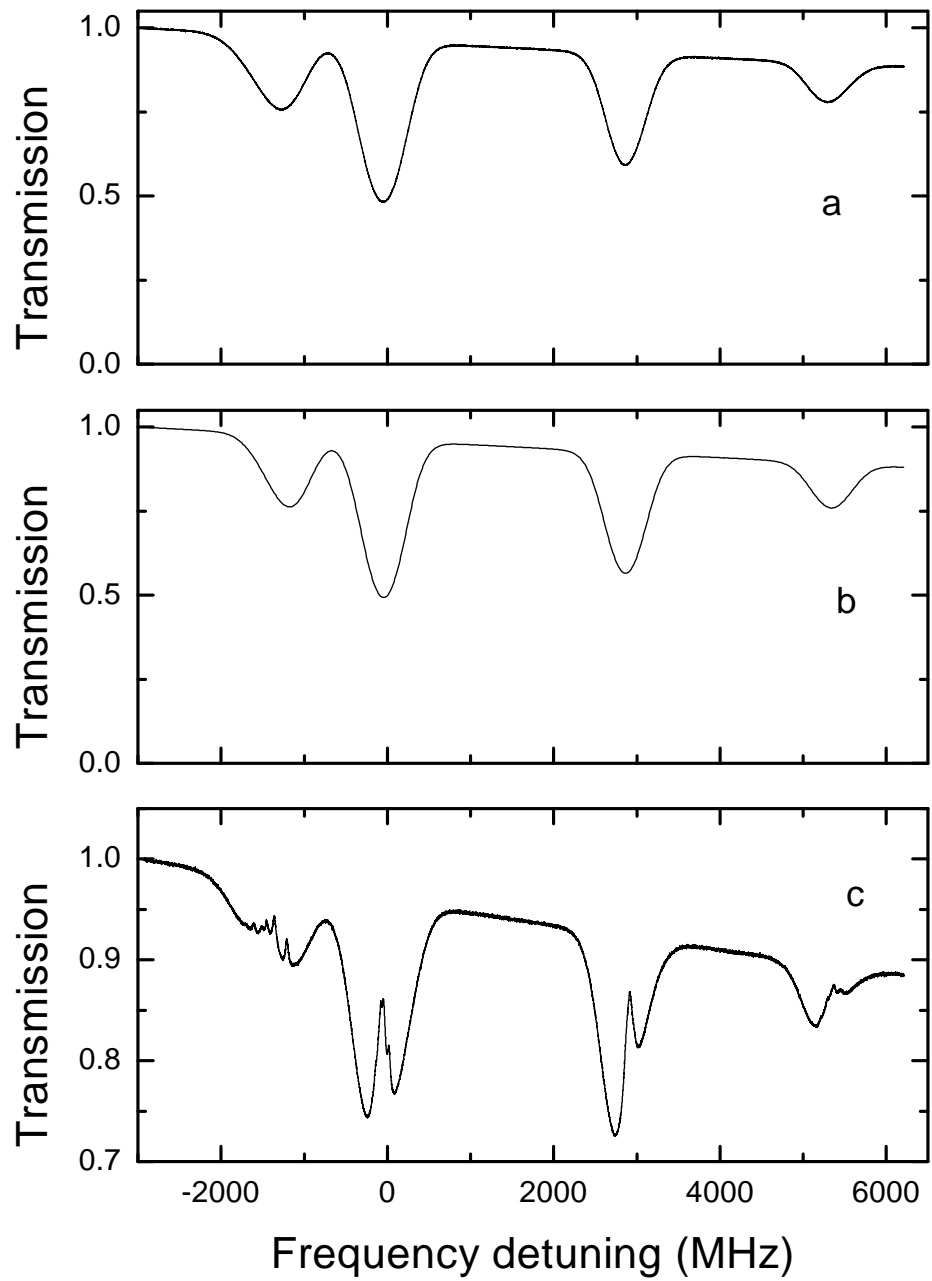}
\caption{Hyperfine spectra and saturated absorption resonances of the Rb $D_2$ 
line. a - low field transmission spectra; b - calculated low field transmission
spectra; c - saturated absorption resonances.}
\label{fig4}
\end{figure}

Fine frequency of the laser radiation was performed by simultaneous tilting 
the grating and linear modulation of the diode electrical current.
Frequency tuning was tested by measuring the
linear transmission spectra of Rb vapour. For this purpose a weak
laser beam passed through a 7.5~cm cell containing Rb vapour at room
temperature. The cell contained natural mixture of Rb isotopes and 
no buffer gas. (The strong counter propagating
beam (Fig.~\ref{fig3}) was closed in this measurements.) 
The data are presented in Fig.~\ref{fig4}a. This spectrum is the 
average of 16 frequency
scans taken during $\simeq1$~s. This result shows that the
range of the laser frequency fine tuning is larger than 9~GHz.
Small decrease of the laser power visible in the  Fig.~\ref{fig4}a is due to the
diode current modulation mentioned above.

\begin{figure}[htb]
\includegraphics[width=14cm]{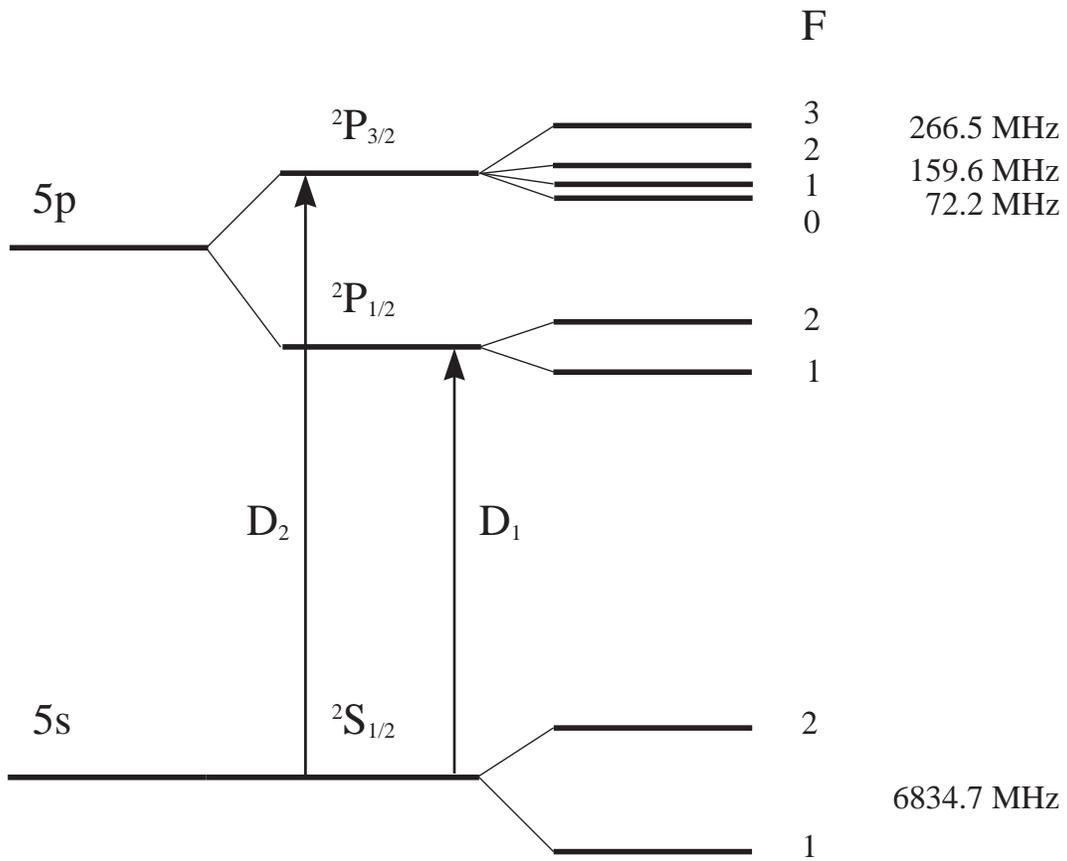}
\caption{Energy levels of the $^{87}$Rb isotope \cite{Barwood91APB}.}
\label{fig5}
\end{figure}

One can compare the measured transmission spectra with 
the calculated spectrum.
The calculation was done using the  hyperfine parameters 
of the $^{87}$Rb and $^{85}$Rb isotopes from
\cite{Barwood91APB} and the decay rate of the $^2P_{3/2}$ state
equal 6~MHz \cite{Schmieder70PRA}.
As an example, the level structure of $^{87}$Rb is shown in Fig.~\ref{fig5}. 
The calculated transmission spectra is shown
in Fig.~\ref{fig4}b.
The measured transmission is reproduced by the calculations within
5\% if one uses the Rb vapour pressure from \cite{Handbook93}. The vapour 
pressure  from \cite{Nesmeyanov63} needs $\simeq$20\% increase. 

Passive stability of the laser frequency was measured by the two methods. In the first,
the transmitted radiation intensity in the vicinity of Rb absorption lines
was observed. The laser frequency drift appeared to be 
$\simeq$12~MHz/min during in approximately 20 min. Another measurement 
was done with the help of the $\lambda$-meter. It gives 
similar value for the frequency drift, 15~MHz/min, for the 20~min 
observation period. 

\begin{figure}[htb]
\includegraphics[width=14cm]{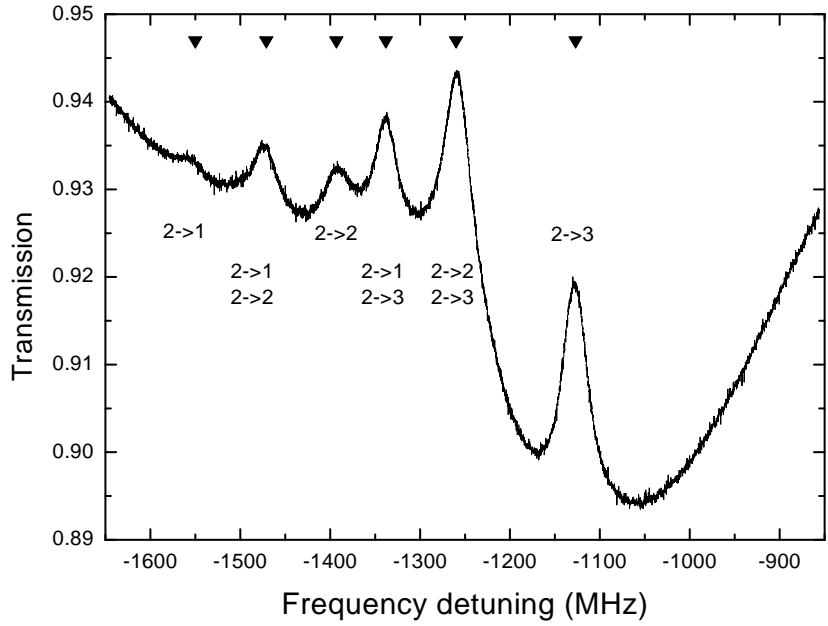}
\caption{Saturated absorption resonances, $F_g=2\rightarrow F_e$ of $^{87}$Rb. 
The main resonances are indicated by one pair of numbers, $F_g=2\rightarrow F_e$; 
the cross-resonances are indicated by two pairs of numbers. Filled triangulars 
indicate the positions of resonances from Ref.~\cite{Barwood91APB}.}
\label{fig6}
\end{figure}

An important parameter of the laser is the radiation linewidth. 
It can be measured studying the beats of two independent lasers. We have
estimated the radiation linewidth using the saturated absorption resonances in
Rb vapour. For this, the transmission of weak laser beam was measured in the
presence of strong counterpropagating laser beam. The spectra are
presented in Fig.~\ref{fig4}c and in a larger scale in Fig.~\ref{fig6}.
These spectra are an average of 16 tracks taken during $\simeq$1~s. 
The width (HWHM) of the nonlinear resonance $F_g=2\longrightarrow F_e=3$
appears to be equal $\simeq15$~MHz.  This width is composed \cite{Rautian79} 
from the fluorescence linewidth (3~MHz for our transition \cite{Schmieder70PRA}), 
the width of the Bennett peak in the velocity distribution in the upper 
state, estimated as 11~MHz for the strong beam
intensity used in the experiment and the laser frequency linewidth. 
From these data one can
estimate that the laser radiation linewidth is on the order of 1 ~MHz.  

\section{Conclusions}

The diode laser having V-shape external cavity is described. The external cavity is 
a Littrow resonator modified by an additional output coupling mirror.
This mirror allows easily adjust the level of the laser output coupling
and thus optimize the laser performance, obtaining necessary frequency
tuning range, or necessary output power. In the described laser the
radiation power constitutes $\simeq$50\% of the laser power without 
an external resonator. The fine frequency tuning range of the laser 
is $\geq$9~GHz and the radiation linewidth is estimated as being
on the order of 1~MHz.

Proposed V-shape external resonator has some advantages in comparison
with the traditional Littrow resonator. It is easier now to optimize
the diffraction grating that has to have just maximal possible reflection
coefficient. Another advantage is that the direction of the main output 
beam is unaffected by the laser frequency tuning.   

\section*{Acknowledgments}

The work was supported by the Russian Foundation for Basic Research
(grant RFBR 03-02-17553) and by the Siberian Branch of the Russian 
Academy of Sciences through the project "Laser cooling of gases in 
magneto-optical traps".


\newpage

\newpage

\newpage

\newpage

\newpage

\newpage

\end{document}